\documentclass[twocolumn,10pt]{asme2ej}
\usepackage{amssymb}
\usepackage{amsfonts}
\usepackage{amsmath}
\usepackage{euscript}
\usepackage{frcursive}
\usepackage{color}
\usepackage{graphicx}
\usepackage{rotating}
\usepackage{bm}
\usepackage{curves}
\usepackage[svgnames,dvipsnames]{pstricks}
\usepackage{lettrine}
\usepackage{mathrsfs}
\usepackage{booktabs}
\usepackage{scalefnt}
\usepackage[normalem]{ulem}
\usepackage{dcolumn}
\usepackage[numbers,sort&compress]{natbib}
\usepackage{tikz}
\DeclareGraphicsExtensions{{},.pdf}
\providecommand{\tsn}[1]{{\text{\scalefont{0.80}#1}}}

\providecommand{\figref}[1]{{\textup{(Fig.~\ref{#1})}}}

\providecommand{\figrefnp}[1]{{\textup{Fig.~\ref{#1}}}}

\providecommand{\eqrefsatob} [2]{\textup{(\ref{#1}--\ref{#2})}}
\providecommand{\eqrefsab}   [2]{\textup{(\ref{#1}, \ref{#2})}}

\providecommand{\figrefsab}   [2]{{\textup{(Figs.~\ref{#1}, \ref{#2})}}}

\providecommand{\parref}[1]{{\textup{(\S\ref{#1})}}}

\providecommand{\ie}{{\em ie~}}

\providecommand{\viz}{{\em viz~}}
\newcommand{\tsr}[1]{{\pmb{\mathsf{#1}}}}

\newcommand{\tr}{\mathrm{tr}}
   \newcommand{\I}[1]{  \mathrm{I}_{\pmb{\mathsf{#1}}}}
  \newcommand{\II}[1]{ \mathrm{II}_{\pmb{\mathsf{#1}}}}
 \newcommand{\III}[1]{\mathrm{III}_{\pmb{\mathsf{#1}}}}
\providecommand{\TKE}{{\mathrm{k}}}
\def\NumERICCS{N\kern-.09em\lower.5ex\hbox{\tsn{um}}\kern-.09em\tsn{ERICCS}}
\newcommand{\ac}{a\mkern-2.5mu c}
\title{Algebraic proof and application of Lumley's realizability triangle}
\author{G.A. Gerolymos and I. Vallet
        \affiliation{Sorbonne Universit\'es, Universit\'e Pierre-et-Marie-Curie (\tsn{UPMC}), 4 place Jussieu, 75005 Paris, France\\
                     Emails: georges.gerolymos@upmc.fr, isabelle.vallet@upmc.fr%; \today
                    }
       }
\begin{document}
\maketitle
%-----------------------------------------------------------------------------------------------------------------------------------
\begin{abstract}
Lumley [Lumley J.L.: {\em Adv. Appl. Mech.} {\bf 18} (1978) 123--176] provided a geometrical proof that any Reynolds-stress tensor $\overline{u_i'u_j'}$
(indeed any tensor whose eigenvalues are invariably nonnegative) should remain inside the so-called Lumley's realizability triangle.
An alternative formal algebraic proof is given that the anisotropy invariants of any positive-definite symmetric Cartesian rank-2 tensor in the 3-D
Euclidian space $\mathbb{E}^3$ define a point which lies within the realizability triangle. This general result applies therefore not only to $\overline{u_i'u_j'}$
but also to many other tensors that appear in the analysis and modeling of turbulent flows. Typical examples are presented based on \tsn{DNS} data for plane channel flow.
\end{abstract}
%-----------------------------------------------------------------------------------------------------------------------------------
%
%-----------------------------------------------------------------------------------------------------------------------------------
%
%
%
%
%
%
%
%
%
\section{Introduction}\label{APALRT_s_I}
%
%
%
%
%
%
%
%
%
%-----------------------------------------------------------------------------------------------------------------------------------

The introduction in \cite{Lumley_Newman_1977a} of Lumley's \cite{Lumley_1978a} realizability triangle is without doubt one of the most important contributions to statistical turbulence theory.
The Reynolds-stress tensor property that serves to prove that every possible (realizable) Reynolds-stress tensor should lie within Lumley's \cite{Lumley_1978a} realizability triangle,
is the positivity of the diagonal components of the covariance of velocity-fluctuations
\begin{subequations}
                                                                                                                                    \label{Eq_APALRT_s_I_001}
\begin{alignat}{6}
r_{ij}:=\overline{u_i'u_j'}
                                                                                                                                    \label{Eq_APALRT_s_I_001a}
\end{alignat}
in every reference-frame, and hence also in the frame of its principal axes \cite{Lumley_1978a},
implying that the tensor $\overline{u_i'u_j'}$ is positive-definite \citep[Theorem 2.3, p. 186]{Stewart_1998a}, is exactly the same as that behind Schumann's \cite{Schumann_1977a} realizability conditions.
Throughout the paper, $u_i\in\{u,v,w\}$ are the velocity components in a Cartesian coordinates system $x_i\in\{x,y,z\}$, $\nu$ is the kinematic viscosity, $(\cdot)'$ denotes Reynolds (ensemble) fluctuations and $\overline{(\cdot)}$
denotes Reynolds (ensemble) averaging.

Lee and Reynolds \cite{Lee_Reynolds_1987a} further argued that Lumley's \cite{Lumley_1978a} realizability triangle also applies to the dissipation tensor
\begin{alignat}{6}
\varepsilon_{ij}:=2\nu\overline{\dfrac{\partial u_i'}{\partial x_k}\dfrac{\partial u_j'}{\partial x_k}}
                                                                                                                                    \label{Eq_APALRT_s_I_001b}
\end{alignat}
and to the covariance of the fluctuating vorticity components
\begin{alignat}{6}
\zeta_{ij}:=\overline{\omega_i'\omega_j'}
                                                                                                                                    \label{Eq_APALRT_s_I_001c}
\end{alignat}
\end{subequations}
where $\omega_i'$ are the fluctuating vorticity components. Obviously the diagonal components of both these tensors are positive for every orientation of the axes of the Cartesian coordinates system.

Realizability constraints are essential not only in theory and modelling \cite{Lumley_Newman_1977a,
                                                                               Lumley_1978a}
but also in computational implementations of second-moment closures \cite{Gerolymos_Vallet_2005a,
                                                                          Gerolymos_Vallet_2009a}.
The same positivity of the diagonal components for every orientation of the axes of coordinates, which is equivalent to the positive-definiteness of the symmetric real tensor \citep[Theorem 2.2, p. 186]{Stewart_1998a},
and implies Lumley's \cite{Lumley_1978a} realizability triangle, can also be of interest to the unresolved stresses \cite{Speziale_1998a} in partially-resolved approaches \cite{Girimaji_2006a}.

Lumley's \cite{Lumley_1978a} proof of the realizability triangle is geometric, based on representing the behaviour of 2 of the principal values of the traceless anisotropy tensor, and taking into account the corresponding behaviour of
the invariants. An alternative easy-to-follow algebraic proof is possible, based on just 2 requirements
\begin{enumerate}
\item the symmetric Reynolds-stress tensor has 3 real eigenvalues \cite[Theorem 2, p. 55]{Segel_2007a}
\item which are nonnegative \cite{Schumann_1977a,Lumley_1978a} with nonzero trace (positive kinetic energy)
\end{enumerate}
which also apply to any symmetric real positive-definite rank-2 tensor in $\mathbb{E}^3$.

%-----------------------------------------------------------------------------------------------------------------------------------
%
%
%
%
%
%
%
%
%
\section{Anisotropy, principal axes and invariants}\label{APALRT_s_APAI}
%
%
%
%
%
%
%
%
%
%-----------------------------------------------------------------------------------------------------------------------------------

\begin{subequations}
                                                                                                                                    \label{Eq_APALRT_s_APAI_001}
Before giving the proof, we summarize for completeness some basic definitions and properties \cite{Lumley_1978a,
                                                                                                   Segel_2007a}.
The tensor of the 2-moments of fluctuating velocities $r_{ij}$ \eqref{Eq_APALRT_s_I_001a} is real and symmetric, and is therefore diagonalizable in the frame of its principal axes \cite[Theorem 5, p. 59]{Segel_2007a},
where its diagonal components (principal values) are its real \cite[Theorem 2, p. 55]{Segel_2007a} eigenvalues \cite[Theorem 4, p. 58]{Segel_2007a}.
This implies that the eigenvalues of $\tsr{r}$, being its diagonal components in the frame of its principal axes, are nonnegative.
Since the eigenvalues of the symmetric tensor $\tsr{r}$ are nonnegative, $\tsr{r}$ is positive-semidefinite \citep[Theorem 2.3, p. 186]{Stewart_1998a}. Inversely, the diagonal components of every positive-semidefinite tensor
are nonnegative \citep[p. 186]{Stewart_1998a}. The halftrace of $\tsr{r}$ is the turbulent kinetic energy and is therefore nonzero ($\tr\tsr{r}=2\TKE>0$), implying that at least one of its eigenvalues is nonzero
(therefore $\tsr{r}$ is positive-definite, which inversely implies nonzero trace).
Let $\tsr{\Lambda_r}$ and $\tsr{\Lambda_{b_r}}$
\begin{alignat}{6}
\tsr{\Lambda_r}:=\begin{bmatrix}\lambda_{r_1}&0&0\\
                                0&\lambda_{r_2}&0\\
                                0&0&\lambda_{r_3}\end{bmatrix}\quad;\quad\tsr{\Lambda_{b_r}}:=\begin{bmatrix}\lambda_{{b_r}_1}&0&0\\
                                                                                                             0&\lambda_{{b_r}_2}&0\\
                                                                                                             0&0&\lambda_{{b_r}_3}\end{bmatrix}
                                                                                                                                    \label{Eq_APALRT_s_APAI_001a}
\end{alignat}
be the diagonal matrices of the eigenvalues of $\tsr{r}$ and $\tsr{b_r}$, respectively, where 
\begin{alignat}{6}
\tsr{b_r}:=\dfrac{\tsr{r}}{\tr\tsr{r}}-\tfrac{1}{3}\tsr{I}_3\implies\left\{\begin{array}{lcl}\I{b_r}  &=&\tr\tsr{b_r}=0\\
                                                                                             \II{b_r} &=&-\tfrac{1}{2}b_{r_{ij}}b_{r_{ji}}\\
                                                                                                      &=&-\tfrac{1}{2}\lambda_{{b_r}_i}\lambda_{{b_r}_i}<0\\
                                                                                             \III{b_r}&=&\mathrm{det}\tsr{b_r}=\lambda_{{b_r}_1}\lambda_{{b_r}_2}\lambda_{{b_r}_3}\\\end{array}\right.
                                                                                                                                    \label{Eq_APALRT_s_APAI_001b}
\end{alignat}
\end{subequations}
is the traceless anisotropy tensor corresponding to $\tsr{r}$ and $\tsr{I}_3$ is the $3\times3$ identity tensor, with the usual definition of the invariants \cite[(6), p. 51]{Segel_2007a},
simplified \cite{Lumley_1978a} by the relation $\I{b_r}=\tr\tsr{b_r}=0$ \eqref{Eq_APALRT_s_APAI_001b}, and the corresponding expressions in terms of the eigenvalues in the frame of principal axes \cite{Lumley_1978a}.
By definition \eqref{Eq_APALRT_s_APAI_001b}, $\tsr{b_r}$ is real symmetric, and has therefore real eigenvalues \cite[Theorem 2, p. 55]{Segel_2007a}.
It is straightforward to show that the eigenvectors of $\tsr{r}$ are also eigenvectors of $\tsr{b_r}$.
Let $\tsr{X_r}$ and $\tsr{X_{b_r}}$ denote the orthonormal matrices \cite[Theorem 5, p. 59]{Segel_2007a} whose columns are the right eigenvectors of $\tsr{r}$ and $\tsr{b_r}$, respectively,
and therefore satisfy
\begin{subequations}
                                                                                                                                    \label{Eq_APALRT_s_APAI_002}
\begin{alignat}{6}
\tsr{r}\cdot\tsr{X_r}=\tsr{\Lambda_r}\cdot\tsr{X_r}\quad;\quad\tsr{b_r}\cdot\tsr{X_{b_r}}=\tsr{\Lambda_{b_r}}\cdot\tsr{X_{b_r}}
                                                                                                                                    \label{Eq_APALRT_s_APAI_002a}
\end{alignat}
By straightforward computation using \eqrefsab{Eq_APALRT_s_APAI_001b}
                                              {Eq_APALRT_s_APAI_002a}
\begin{alignat}{6}
\tsr{b_r}\cdot\tsr{X_r}\stackrel{\eqref{Eq_APALRT_s_APAI_001b}}{=}&\left(\dfrac{\tsr{r}}{\tr\tsr{r}}-\tfrac{1}{3}\tsr{I}_3\right)\cdot\tsr{X_r}
                                                               =  \dfrac{1}{\tr\tsr{r}}\tsr{r}\cdot\tsr{X_r}-\tfrac{1}{3}\tsr{X_r}
                                                                                                                                    \notag\\
                       \stackrel{\eqref{Eq_APALRT_s_APAI_002a}}{=}&\left(\dfrac{1}{\tr\tsr{r}}\tsr{\Lambda_r}-\tfrac{1}{3}\tsr{I}_3\right)\cdot\tsr{X_r}
                                                                                                                                    \label{Eq_APALRT_s_APAI_002b}
\end{alignat}
implying by \eqref{Eq_APALRT_s_APAI_002a}
\begin{alignat}{6}
\tsr{X_{b_r}}      \stackrel{\eqrefsab{Eq_APALRT_s_APAI_002a}
                                      {Eq_APALRT_s_APAI_002b}}{=}\tsr{X_r}
                                                                                                                                    \label{Eq_APALRT_s_APAI_002c}\\
\tsr{\Lambda_{b_r}}\stackrel{\eqrefsab{Eq_APALRT_s_APAI_002a}
                                      {Eq_APALRT_s_APAI_002b}}{=}\dfrac{1}{\tr\tsr{r}}\tsr{\Lambda_r}-\tfrac{1}{3}\tsr{I}_3
\iff\lambda_{r_i}=\left(\tr\tsr{r}\right)\left(\lambda_{b_{r_i}}+\tfrac{1}{3}\right)
                                                                                                                                    \label{Eq_APALRT_s_APAI_002d}
\end{alignat}
\ie\ $\tsr{r}$ and $\tsr{b_r}$ have the same system of principal axes and their eigenvalues are related by \eqref{Eq_APALRT_s_APAI_002d}.
\end{subequations}

%-----------------------------------------------------------------------------------------------------------------------------------
%
%
%
%
%
%
%
%
%
\section{Proof}\label{APALRT_s_P}
%
%
%
%
%
%
%
%
%
%-----------------------------------------------------------------------------------------------------------------------------------

\begin{figure*}[ht]
\begin{center}
\begin{picture}(450,350)
\put(-10,0){\includegraphics[width=450pt,bb=63 377 570 750]{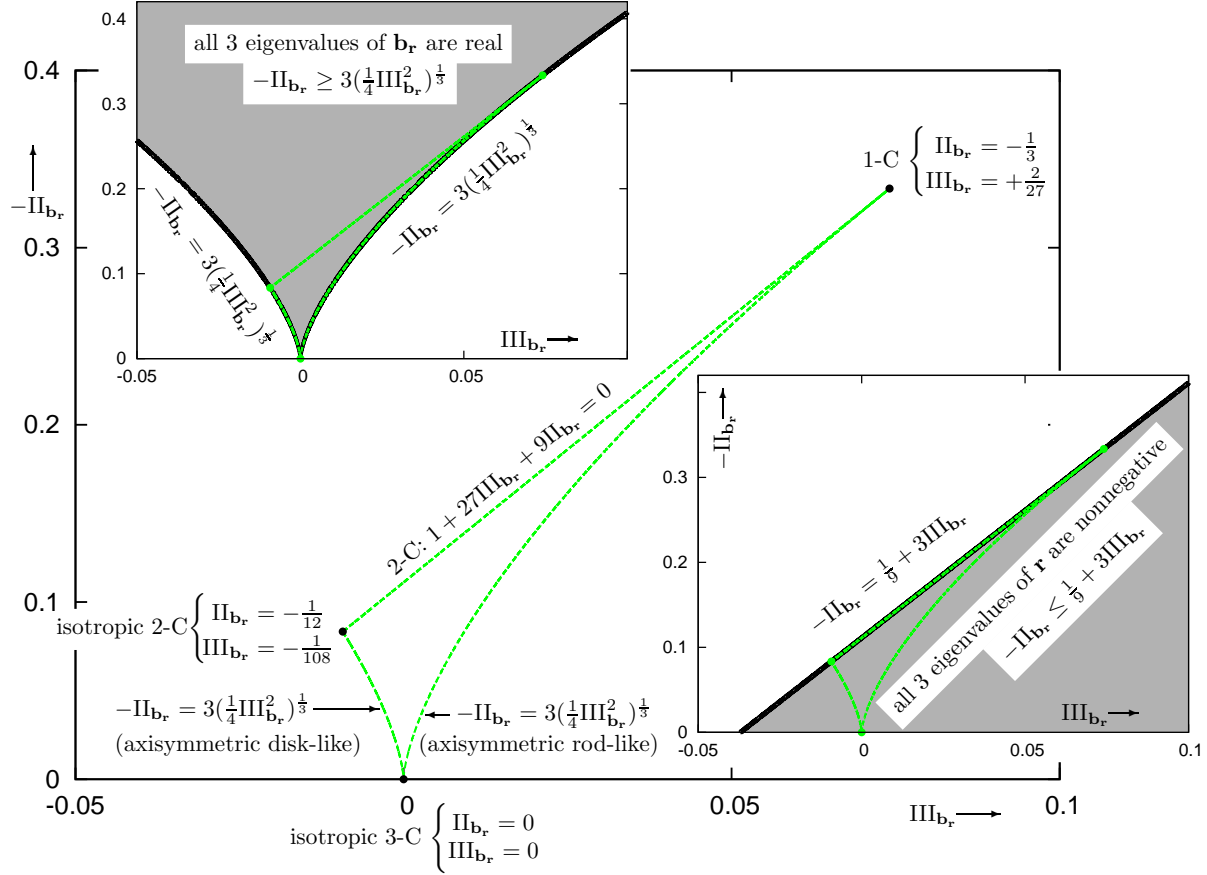}}
\end{picture}
\end{center}
\caption{Lumley's \cite{Lumley_1978a,
                        Simonsen_Krogstad_2005a} realizability triangle \eqref{Eq_APALRT_s_P_003} for a positive-definite symmetric real rank-2 Cartesian tensor $\tsr{r}$ in the 3-D Euclidean space,
plotted in the $(\III{b_r},-\II{b_r})$-plane of the invariants \eqref{Eq_APALRT_s_APAI_001b} of the corresponding anisotropy tensor $\tsr{b_r}$ \eqref{Eq_APALRT_s_APAI_001b},
and representation of the inequalities \eqrefsab{Eq_APALRT_s_P_001b}
                                                {Eq_APALRT_s_P_002} whose intersection defines the region of realizable states.}
\label{Fig_APALRT_s_P_001}
\end{figure*}

As stated in the introduction the algebraic proof of Lumley's \cite{Lumley_1978a} realizability triangle can be easily obtained from 2 well-known conditions, also discussed in \parref{APALRT_s_APAI}.
The eigenvalues of $\tsr{b_r}$ satisfy the characteristic polynomial \cite[(5), p. 51]{Segel_2007a}
\begin{subequations}
                                                                                                                                    \label{Eq_APALRT_s_P_001}
\begin{alignat}{6}
\lambda_{b_r}^3-\underbrace{\I{b_r}}_{\stackrel{\eqref{Eq_APALRT_s_APAI_001b}}{=}0}\,\lambda_{b_r}^2+\II{b_r}\,\lambda_{b_r}-\III{b_r}=0
                                                                                                                                    \label{Eq_APALRT_s_P_001a}
\end{alignat}
The roots of the cubic equation \eqref{Eq_APALRT_s_P_001a} are real iff \cite[pp. 44--45]{Harris_Stocker_1998a}
its determinant is nonpositive,\footnote{\label{ff_APALRT_s_P_001}The cubic equation $x^3+ax^2+bx+c=0$ has 3 real roots iff \cite[pp. 44--45]{Harris_Stocker_1998a}
                                                                  the determinant is negative, $\Delta_3=[\tfrac{1}{9}(3b-a^2)]^3+[\tfrac{1}{2}(c+\tfrac{2}{27}a^3-\tfrac{2}{27}ab)]^2\leq0$.
                                                                  If $\Delta_3=0$ then 2 roots are identical.}
\ie
\begin{eqnarray}
\text{$\tsr{b_r}$ has 3 real eigenvalues}\iff&\left(\tfrac{1}{9}(3\II{b_r})\right)^3+\left(\tfrac{1}{2}\III{b_r}\right)^2\leq0
                                                                                                                                    \notag\\
                                         \iff&-\II{b_r}\geq3\left(\tfrac{1}{4}\III{b_r}^2\right)^\frac{1}{3}
                                                                                                                                    \label{Eq_APALRT_s_P_001b}
\end{eqnarray}
\end{subequations}
readily implying that possible (realizable) states must lie \figref{Fig_APALRT_s_P_001} above the 2 branches of axisymmetric componentality in the $(\III{b_r},-\II{b_r})$-plane \cite{Simonsen_Krogstad_2005a}.
Furthermore, the eigenvalues of $\tsr{r}$ representing also its diagonal components in the system of principal axes \cite[Theorem 4, p. 58]{Segel_2007a} must be nonegative,
also implying that $\mathrm{det}\tsr{r}=\lambda_{r_1}\lambda_{r_2}\lambda_{r_3}\geq0$, the last of the 3 realizability conditions of Shumann \cite{Schumann_1977a}.
Using the relation \eqref{Eq_APALRT_s_APAI_002d} between the eigenvalues of $\tsr{r}$ and $\tsr{b_r}$
\begin{align}
0 \leq&\dfrac{1}{\left(\tr\tsr{r}\right)^3}\lambda_{r_1}\lambda_{r_2}\lambda_{r_3}
  \stackrel{\eqref{Eq_APALRT_s_APAI_002d}}{=} \left(\lambda_{b_{r_1}}+\tfrac{1}{3}\right)\left(\lambda_{b_{r_2}}+\tfrac{1}{3}\right)\left(\lambda_{b_{r_3}}+\tfrac{1}{3}\right)
                                                                                                                                    \notag\\
  \stackrel{\eqref{Eq_APALRT_s_APAI_001}}{=}&\tfrac{1}{27}+\tfrac{1}{3}\II{b_r}+\III{b_r}\iff-\II{b_r}\leq\tfrac{1}{9}+3\III{b_r}
                                                                                                                                    \label{Eq_APALRT_s_P_002}
\end{align}
implying that possible (realizable) states equally lie \figref{Fig_APALRT_s_P_001} below the 2-C straight line in the $(\III{b_r},-\II{b_r})$-plane \cite{Simonsen_Krogstad_2005a}.
The intersection of these 2 inequalities is precisely Lumley's \cite{Lumley_1978a} realizability triangle
\begin{eqnarray}
\eqrefsab{Eq_APALRT_s_P_001b}
         {Eq_APALRT_s_P_002} \implies
3\left(\tfrac{1}{4}\III{b_r}^2\right)^\frac{1}{3}\leq-\II{b_r}\leq\tfrac{1}{9}+3\III{b_r}
                                                                                                                                    \label{Eq_APALRT_s_P_003}
\end{eqnarray}
and is defined by the 2 conditions stated in \parref{APALRT_s_I}, \viz\ that the eigenvalues of $\tsr{r}$ are real and positive.

Notice that \eqref{Eq_APALRT_s_P_003} describes precisely a curvilinear triangle \figref{Fig_APALRT_s_P_001}, because the 2 axisymmetric branches
$-\II{b_r}=3\left(\tfrac{1}{4}\III{b_r}^2\right)^\frac{1}{3}$ \eqref{Eq_APALRT_s_P_001b} are obviously described by the same single-valued function of $\III{b_r}$
with a cusp at $(-\II{b_r},\III{b_r})=(0,0)$, defining the isotropic 3-C corner \cite[Fig. 4, p. 3]{Simonsen_Krogstad_2005a}.
The intersections of this single valued function with the straight line $-\II{b_r}=\tfrac{1}{9}+3\III{b_r}$ \eqref{Eq_APALRT_s_P_002}
are the roots of $3\left(\tfrac{1}{4}\III{b_r}^2\right)^\frac{1}{3}=\tfrac{1}{9}+3\III{b_r}\iff
                   \III{b_r}^3-\tfrac{5}{36}\III{b_r}^2+\tfrac{1}{243}\III{b_r}+\tfrac{1}{19683}=0\iff
                   \left(\III{b_r}+\tfrac{1}{108}\right)\left(\III{b_r}-\tfrac{2}{27}\right)^2=0$,
defining the 2 other corners, \viz\ the 1-C corner \cite[Fig. 4, p. 3]{Simonsen_Krogstad_2005a} $(-\II{b_r},\III{b_r})=(\tfrac{1}{3},\tfrac{2}{27})$
corresponding to the double root $\tfrac{2}{27}$ and the isotropic 2-C point \cite[Fig. 4, p. 3]{Simonsen_Krogstad_2005a} $(-\II{b_r},\III{b_r})=(\tfrac{1}{12},-\tfrac{1}{108})$.
\begin{figure*}[ht]
\begin{center}
\begin{picture}(450,580)
\put( 30,-15){\includegraphics[width=380pt,bb=56 51 515 763]{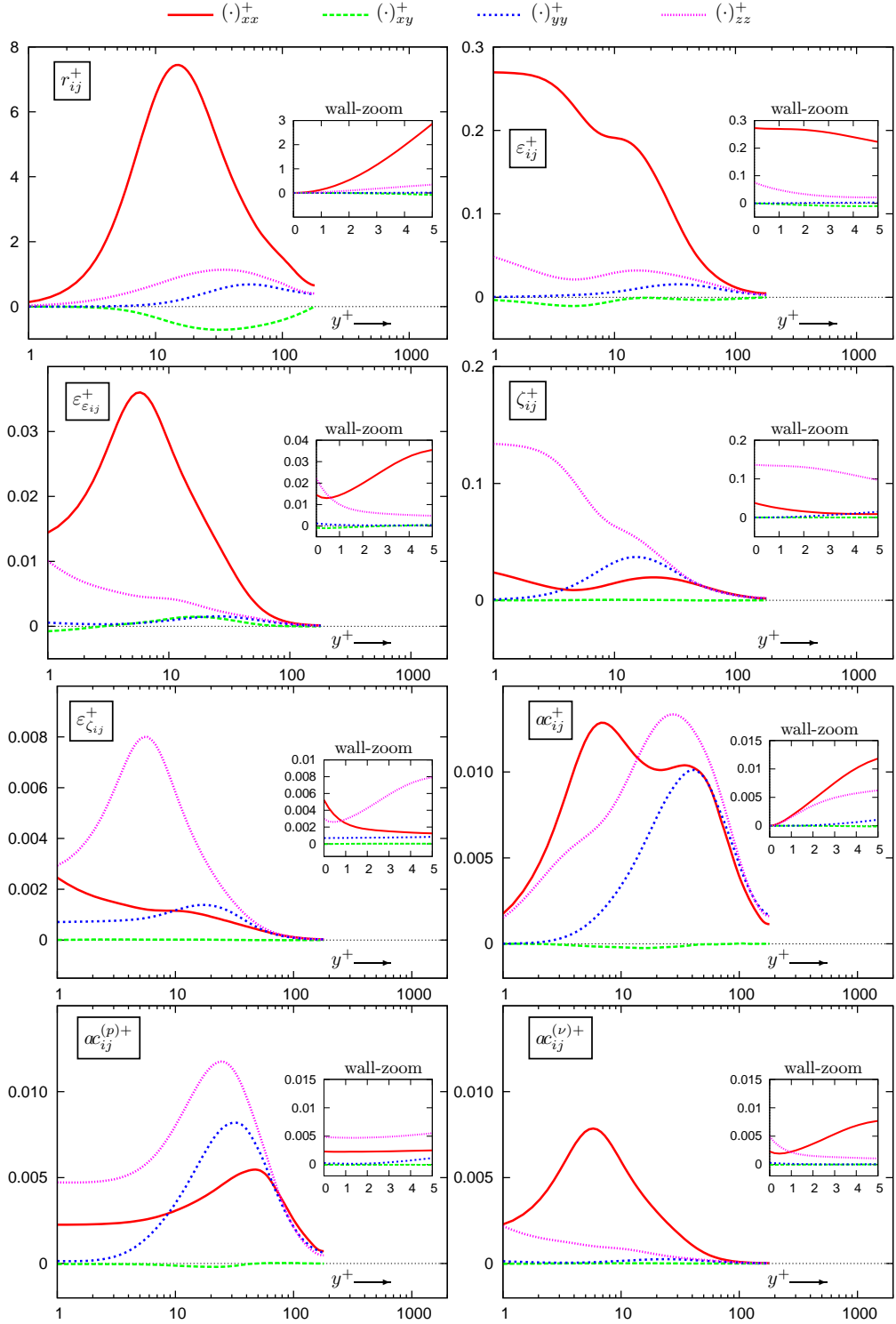}}
\end{picture}
\end{center}
\caption{Components, in wall-units \cite[(A3), p. 28]{Gerolymos_Vallet_2016a_news}, of the positive-definite symmetric tensors \eqrefsab{Eq_APALRT_s_I_001}
                                                                                                                                         {Eq_APALRT_s_A_001},
plotted against the inner-scaled wall-distance $y^+$ (logscale and linear wall-zoom),
from \tsn{DNS} computations of turbulent plane channel flow \cite{Gerolymos_Vallet_2016a_news,
                                                                  Gerolymos_Vallet_2016b_news}.}
\label{Fig_APALRT_s_A_001}
\end{figure*}
\begin{figure*}[ht]
\begin{center}
\begin{picture}(450,580)
\put(-15,-15){\includegraphics[width=480pt,bb=43 52 597 738]{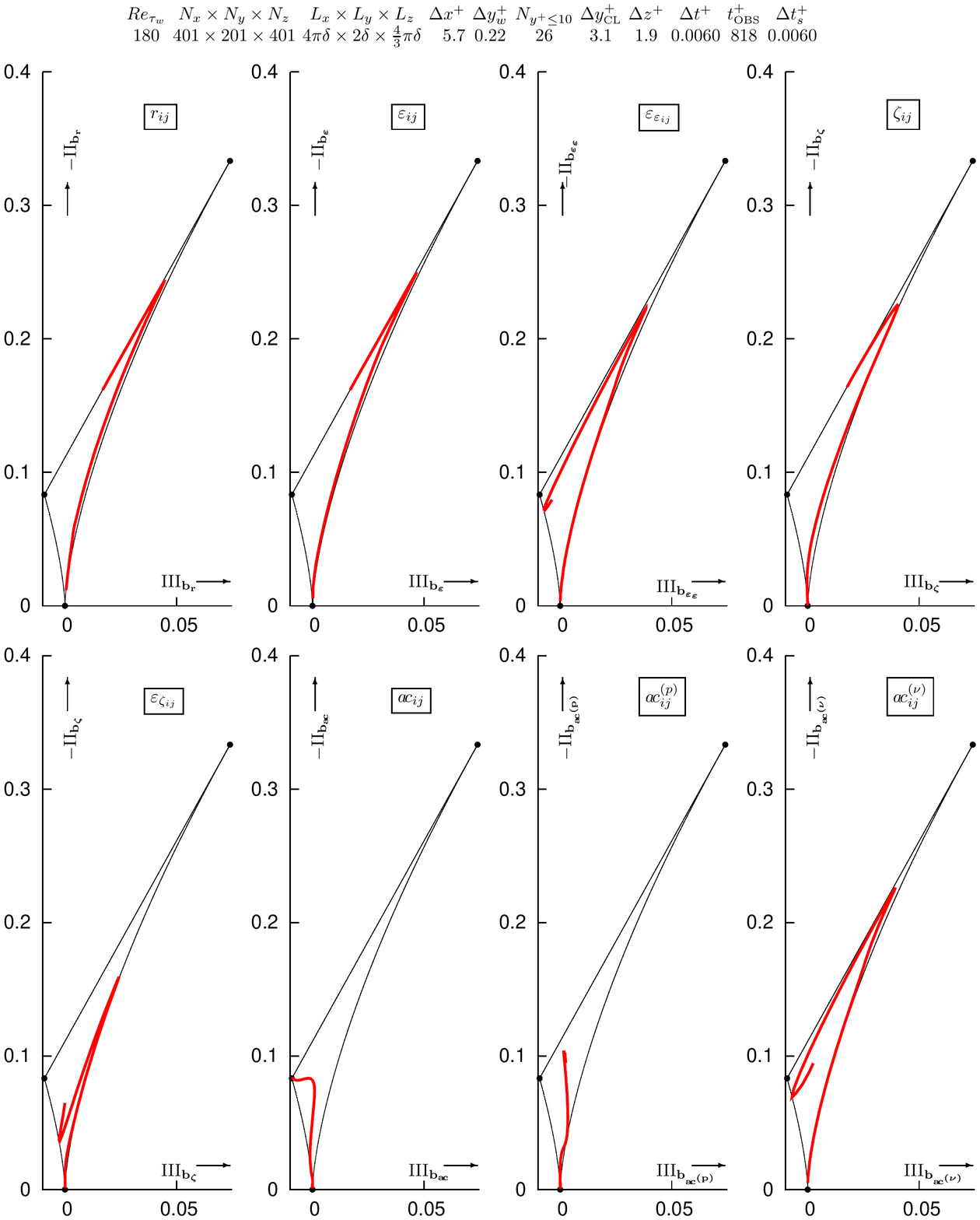}}
\end{picture}
\end{center}
\caption{Locus, within Lumley's \cite{Lumley_1978a,
                                      Simonsen_Krogstad_2005a} realizability triangle \eqref{Eq_APALRT_s_P_003} in the $(\III{},-\II{})$-plane,
of the anisotropy invariants \eqref{Eq_APALRT_s_APAI_001b} of the positive-definite symmetric tensors \eqrefsab{Eq_APALRT_s_I_001}
                                                                                                               {Eq_APALRT_s_A_001},
from \tsn{DNS} computations of turbulent plane channel flow \cite{Gerolymos_Vallet_2016a_news,
                                                                  Gerolymos_Vallet_2016b_news}.}
\label{Fig_APALRT_s_A_002}
\end{figure*}

%-----------------------------------------------------------------------------------------------------------------------------------
%
%
%
%
%
%
%
%
%
\section{Applications}\label{APALRT_s_A}
%
%
%
%
%
%
%
%
%
%-----------------------------------------------------------------------------------------------------------------------------------

Obviously the property applies not only to the Reynolds-stresses $r_{ij}$ \eqref{Eq_APALRT_s_I_001a},
their dissipation $\varepsilon_{ij}$ \eqref{Eq_APALRT_s_I_001b} or the vorticity covariance $\zeta_{ij}$ \eqref{Eq_APALRT_s_I_001c},
but also to any tensor with nonnegative diagonal values.
Typical examples are the destruction-of-dissipation tensor \cite{Gerolymos_Vallet_2016a_news,
                                                                 Gerolymos_Vallet_2016b_news}
\begin{subequations}
                                                                                                                                    \label{Eq_APALRT_s_A_001}
\begin{alignat}{6}
\varepsilon_{\varepsilon_{ij}}:=&4\nu^2\overline{\dfrac{\partial^2 u_i'}{\partial x_k\partial x_m}\dfrac{\partial^2 u_j'}{\partial x_k\partial x_m}}
                                                                                                                                    \label{Eq_APALRT_s_A_001a}
\end{alignat}
which represents the destruction of $\varepsilon_{ij}$ by the action of molecular viscosity \cite[(3.3), p. 17]{Gerolymos_Vallet_2016a_news}
or the destruction-of-vorticity-covariance tensor
\begin{alignat}{6}
\varepsilon_{\zeta{ij}}:=&2\nu\overline{\dfrac{\partial\omega_i'}{\partial x_k}\dfrac{\partial\omega_j'}{\partial x_k}}
                                                                                                                                    \label{Eq_APALRT_s_A_001b}
\end{alignat}
which represents the destruction of $\zeta_{ij}$ by the action of molecular viscosity \cite[(20), p. 458]{Bernard_Berger_1982a}.

Regarding acceleration fluctuations $(D_t u_i)'$ most authors generally study the variances of its components \cite{Yeung_Pope_Lamorgese_Donzis_2006a,
                                                                                                                    Yeo_Kim_Lee_2010a}
and its splitting, based on the momentum equation, in a pressure part $\rho^{-1}\partial_{x_i}p'$ (also called {\em inviscid}) and a viscous part $\nu\nabla^2u_i'$
(also called {\em soleneidal} because, by the fluctuating continuity equation \cite[(3.2a), p. 17]{Gerolymos_Vallet_2016a_news}, it is divergence-free). As for the fluctuating vorticity correlations, we may define the symmetric
positive-definite tensor of fluctuating acceleration correlations $\ac_{ij}$ and the corresponding inviscid and solenoidal parts
\begin{alignat}{6}
\ac_{ij}:=&\overline{\left(\dfrac{Du_i}{Dt}\right)'\left(\dfrac{Du_j}{Dt}\right)'}
                                                                                                                                    \label{Eq_APALRT_s_A_001c}\\
\ac^{(p)}_{ij}:=&\dfrac{1}{\rho^2}\overline{\dfrac{\partial p'}{\partial x_i}\dfrac{\partial p'}{\partial x_j}}
                                                                                                                                    \label{Eq_APALRT_s_A_001d}\\
\ac^{(\nu)}_{ij}:=&\nu^2\overline{\dfrac{\partial^2 u_i'}{\partial x_m\partial x_m}\dfrac{\partial^2 u_j'}{\partial x_k\partial x_k}}
                                                                                                                                    \label{Eq_APALRT_s_A_001e}
\end{alignat}
\end{subequations}

We consider \tsn{DNS} results of fully developed (streamwise invariant in the mean) turbulent plane channel flow \cite{Gerolymos_Vallet_2016a_news,
                                                                                                                       Gerolymos_Vallet_2016b_news},
in a streamwise$\times$wall-normal$\times$spanwise $L_x\times L_y\times L_z=4\pi\delta\times2\delta\times\tfrac{4}{3}\pi\delta$ computational box,
and use standard definitions \cite[\S3.2, p. 18]{Gerolymos_Vallet_2016a_news} of computational parameters \figrefsab{Fig_APALRT_s_A_001}
                                                                                                                     {Fig_APALRT_s_A_002}.
Regarding $r_{ij}$ \eqref{Eq_APALRT_s_I_001a}, its dissipation-rate $\varepsilon_{ij}$ \eqref{Eq_APALRT_s_I_001b} and the destruction of that dissipation $\varepsilon_{\varepsilon_{ij}}$ \eqref{Eq_APALRT_s_A_001a},
notice that the shear component $(\cdot)_{xy}$ is generally of the order-of-magnitude of the wall-normal component $(\cdot)_{yy}$ \figref{Fig_APALRT_s_A_001}.
Sufficiently far from the wall \cite{Lee_Reynolds_1987a} $r_{ij}$ is expected to reflect the anisotropy of the large turbulent scales (typical size $\ell_\tsn{T}$), whereas
$\varepsilon_{ij}$ is expected to reflect the anisotropy of the smaller scales (of the order of the Taylor microscale $\lambda$). The scaling arguments of Tennekes and Lumley \cite[pp. 88--92]{Tennekes_Lumley_1972a}
suggest that, again sufficiently far from the wall, $\varepsilon_{\varepsilon_{ij}}$ reflects the anisotropy of scales between $\lambda$ and the Kolmogorov scale $\ell_\tsn{K}$.
It is therefore noteworthy that they appear to share a seemingly similar anisotropy $(\cdot)_{xx}>(\cdot)_{zz}>(\cdot)_{yy}\;\forall y^+\gtrapprox1$ \figref{Fig_APALRT_s_A_001}.
Nonetheless, very near the wall ($y^+\lessapprox1$; \figrefnp{Fig_APALRT_s_A_001}), where all these lengthscales collapse to $0$,
$\varepsilon_{\varepsilon_{zz}}$ becomes larger than $\varepsilon_{\varepsilon_{xx}}$.
Vorticity covariance $\zeta_{ij}$ \eqref{Eq_APALRT_s_I_001c} is expected \cite[pp. 88--92]{Tennekes_Lumley_1972a} to reflect the anisotropy of the same scales as $\varepsilon_{ij}$,
and its destruction $\varepsilon_{\zeta{ij}}$ \eqref{Eq_APALRT_s_I_001b} corresponding to the same scales as $\varepsilon_{\varepsilon_{ij}}$.
Both $\zeta_{ij}$ and $\varepsilon_{\zeta{ij}}$ have a very weak shear component $(\cdot)_{xy}$ \figref{Fig_APALRT_s_A_001}.
Their componentality obviously differs from that of $\{r_{ij},\varepsilon_{ij},\varepsilon_{\varepsilon_{ij}}\}$, because in the major part of the channel $(\cdot)_{xx}\approxeq(\cdot)_{yy}<(\cdot)_{zz}$
($y^+\gtrapprox10$; \figrefnp{Fig_APALRT_s_A_001}). Nonetheless, $\zeta_{yy}\underset{y^+\to0}{\to}0$ (2-C at the wall), contrary to $\varepsilon_{\zeta{yy}}$ \figref{Fig_APALRT_s_A_001},
and in the sublayer $\varepsilon_{\zeta{xx}}$ and $\varepsilon_{\zeta{zz}}$ cross each other ($y^+\lessapprox1$; \figrefnp{Fig_APALRT_s_A_001}), in analogy with the observed behaviour of $\varepsilon_{\varepsilon_{ij}}$.

Regarding the acceleration correlations, $\ac_{ij}$ \eqref{Eq_APALRT_s_A_001c}, $\ac^{(p)}_{ij}$ \eqref{Eq_APALRT_s_A_001d} and $\ac^{(\nu)}_{ij}$ \eqref{Eq_APALRT_s_A_001e}, again the shear component is
substantially smaller than the diagonal components \figref{Fig_APALRT_s_A_001}. Recall that the fluctuating momentum equation \cite[(3.2b), p. 17]{Gerolymos_Vallet_2016a_news}
\begin{subequations}
                                                                                                                                    \label{Eq_APALRT_s_A_002}
\begin{alignat}{6}
\dfrac{D u_i'}{Dt}=-\dfrac{1}{\rho}\dfrac{\partial p'}{\partial x_i}+\nu\dfrac{\partial^2 u_i'}{\partial x_m\partial x_m}
                                                                                                                                    \label{Eq_APALRT_s_A_002a}
\end{alignat}
readily implies
\begin{alignat}{6}
\eqrefsatob{Eq_APALRT_s_A_001c}
           {Eq_APALRT_s_A_002a}\implies\ac_{ij}=&\ac^{(p)}_{ij}+\ac^{(\nu)}_{ij}
                                                                                                                                    \notag\\
                                               -&\dfrac{\nu}{\rho}\left(\overline{\dfrac{\partial p'}{\partial x_i}\dfrac{\partial^2 u_j'}{\partial x_k\partial x_k}
                                                                                 +\dfrac{\partial p'}{\partial x_j}\dfrac{\partial^2 u_i'}{\partial x_m\partial x_m}}\right)
                                                                                                                                    \label{Eq_APALRT_s_A_002b}
\end{alignat}
where the last cross-correlation tensor is symmetric but indefinite.
The componentality of the acceleration correlations $\ac_{ij}$ \eqref{Eq_APALRT_s_A_001c} is quite different from that
of its pressure $\ac^{(p)}_{ij}$ \eqref{Eq_APALRT_s_A_001d} and viscous $\ac^{(\nu)}_{ij}$ \eqref{Eq_APALRT_s_A_001e} parts,
as these correlations are the footprint of different mechanisms occurring mainly at different scales.
In the buffer layer ($10\lessapprox y^+\lessapprox100$; \figrefnp{Fig_APALRT_s_A_001}) viscous acceleration is mainly in the streamwise direction,
but in the sublayer $\ac^{(\nu)}_{zz}$ increases and crosses with $\ac^{(\nu)}_{xx}$ at $y^+\approxeq1$ \figref{Fig_APALRT_s_A_001},
in analogy with the other correlations between components of the fluctuating velocity Hessian, 
$\varepsilon_{\varepsilon_{ij}}$ \eqref{Eq_APALRT_s_A_001a} and $\varepsilon_{\zeta{ij}}$ \eqref{Eq_APALRT_s_I_001b}.
The wall normal component $\ac^{(\nu)}_{yy}$ becomes comparable to the other diagonal components only sufficiently away
from the wall ($y^+\gtrapprox30$; \figrefnp{Fig_APALRT_s_A_001}).
On the other hand, acceleration induced by fluctuating pressure forces $\ac^{(p)}_{ij}$ \eqref{Eq_APALRT_s_A_001d}
exhibits a $\ac^{(p)}_{zz}>\ac^{(p)}_{yy}>\ac^{(p)}_{xx}$ anisotropy in the buffer layer ($10\lessapprox y^+\lessapprox100$; \figrefnp{Fig_APALRT_s_A_001}),
whereas near the wall $\ac^{(p)}_{yy}\underset{y^+\to0}{\to}0$ ($y^+\lessapprox10$; \figrefnp{Fig_APALRT_s_A_001}).
Finally, the acceleration correlations behave quite differently from the 2 parts in the fluctuating momentum equation \eqref{Eq_APALRT_s_A_002a},
implying that the cross-term in \eqref{Eq_APALRT_s_A_002b} is important, and especially so near the wall where scale-separation tends to disappear,
and is directly responsible for the differences in limiting behavior \figref{Fig_APALRT_s_A_001}
\begin{alignat}{6}
\underset{y^+\to0}{\lim}\ac_{ij}=&0
                                                                                                                                    \label{Eq_APALRT_s_A_002c}\\
\underset{y^+\to0}{\lim}\ac^{(p)}_{ij}\neq&0
                                                                                                                                    \label{Eq_APALRT_s_A_002d}\\
\underset{y^+\to0}{\lim}\ac^{(\nu)}_{ij}\neq&0
                                                                                                                                    \label{Eq_APALRT_s_A_002e}
\end{alignat}
\end{subequations}

More precise information on the componentality of these positive-definite symmetric tensors \eqrefsab{Eq_APALRT_s_I_001}
                                                                                                     {Eq_APALRT_s_A_001}
is obtained by considering their anisotropy invariant mapping (\tsn{AIM}) in the $(\III{},-\II{})$-plane \figref{Fig_APALRT_s_A_002}.
Only $r_{ij}$ \eqref{Eq_APALRT_s_I_001a}, $\varepsilon_{ij}$ \eqref{Eq_APALRT_s_I_001b}, $\zeta_{ij}$ \eqref{Eq_APALRT_s_I_001c} and $\ac_{ij}$ \eqref{Eq_APALRT_s_A_001c}
are 2-C at the wall \figref{Fig_APALRT_s_A_002}.
The fluctuating acceleration correlations $\ac_{ij}$ \eqref{Eq_APALRT_s_A_001c} reach the 2-C state near, although not exactly at, the axisymmetric disk-like boundary \figref{Fig_APALRT_s_A_002}.
The tensors representing correlations between components of the fluctuating velocity Hessian,
$\varepsilon_{\varepsilon_{ij}}$ \eqref{Eq_APALRT_s_A_001a}, $\varepsilon_{\zeta{ij}}$ \eqref{Eq_APALRT_s_I_001b} and $\ac^{(\nu)}_{ij}$ \eqref{Eq_APALRT_s_A_001e},
invariably reach the axisymmetric disk-like boundary of the realizability triangle very near $y^+\approxeq1$ \figref{Fig_APALRT_s_A_002}, roughly where the streamwise $(\cdot)_{xx}$ and
spanwise $(\cdot)_{zz}$ components cross each other \figref{Fig_APALRT_s_A_001},
and then return inside the realizability triangle as they approach the wall \figref{Fig_APALRT_s_A_002}.
Near the centerline, $\ac_{ij}$ approaches disk-like axisymmetry \figref{Fig_APALRT_s_A_002}, contrary to
$\ac^{(p)}_{ij}$ \eqref{Eq_APALRT_s_A_001d} and $\ac^{(\nu)}_{ij}$ \eqref{Eq_APALRT_s_A_001e}, both of wich are axisymmetric rod-like \figref{Fig_APALRT_s_A_002}.
The difference is that $\ac_{xx}<\ac_{yy}\approxeq\ac_{zz}\;\forall\,y^+\gtrapprox50$ \figref{Fig_APALRT_s_A_001},
whereas $\ac^{(p)}_{xx}>\ac^{(p)}_{yy}\approxeq\ac^{(p)}_{zz}\;\forall\,y^+\gtrapprox80$ \figref{Fig_APALRT_s_A_001}.
Finally $\ac^{(p)}_{ij}$ approaches the 2-C boundary without reaching it, and then returns inside the realizability triangle \figref{Fig_APALRT_s_A_002}.
Notice that by \eqrefsab{Eq_APALRT_s_A_002a}
                        {Eq_APALRT_s_A_002c}
$[\ac^{(p)}_{ij}]_w=[\ac^{(\nu)}_{ij}]_w$ at the wall.

%-----------------------------------------------------------------------------------------------------------------------------------
%
%
%
%
%
%
%
%
%
\section{Conclusion}\label{APALRT_s_C}
%
%
%
%
%
%
%
%
%
%-----------------------------------------------------------------------------------------------------------------------------------

The simple algebraic proof presented above, can be summarized in the following mathematical proposition:

{\bf Theorem} (Lumley's realizability triangle). Let $\tsr{r}$ be a real rank-2 Cartesian tensor in the {\rm 3-D} Euclidean space $\mathbb{E}^3$. Assume $\tsr{r}$ symmetric and positive definite.
Then the locus of the invariants \eqref{Eq_APALRT_s_APAI_001b} of the corresponding anisotropy tensor $\tsr{b_r}$ \eqref{Eq_APALRT_s_APAI_001b},
in the $(\III{b_r},-\II{b_r})$-plane, lies within Lumley's realizability triangle \eqref{Eq_APALRT_s_P_003}.\qed
The proof \parref{APALRT_s_P} follows directly from the well-kown fact that the principal values of $\tsr{r}$ are real and nonegative.
By its algebraic nature it leads directly to the inequality \eqref{Eq_APALRT_s_P_003}, which defines Lumley's realizability triangle.
It is obtained in the actual $(\III{b_r},-\II{b_r})$-plane, with no need of transformation of the invariants or explicit analysis
of the limit states at the boundaries of the realizability triangle.
The novel algebraic proof reported in the paper helps to better grasp the classic geometric proof given in Lumley \cite{Lumley_1978a}.

Many symmetric tensors with nonnegative diagonal values are encountered in the analysis of turbulent flows.
Several, by no means exhaustive, examples are studied, using \tsn{DNS} data for plane channel flow,
illustrating how anisotropy invariant mapping (\tsn{AIM}) within the realizability triangle can improve our understanding of their componentality behavior.

%-----------------------------------------------------------------------------------------------------------------------------------
%
%
%
%
%
%
%
%
%
\begin{acknowledgment}The authors are listed alphabetically. The present work was partly supported by the \tsn{ANR} project \NumERICCS\ (\tsn{ANR--15--CE06--0009}).\end{acknowledgment}
%
%
%
%
%
%
%
%
%
%-----------------------------------------------------------------------------------------------------------------------------------
%
%-----------------------------------------------------------------------------------------------------------------------------------
%
%
%
%
%
%
%
%
%
%\footnotesize\bibliographystyle{ASME_Journal_1_0_v2/asmems4}\bibliography{Aerodynamics,GV,GV_news}\normalsize
\footnotesize\bibliographystyle{asmems4}\bibliography{Aerodynamics,GV,GV_news}\normalsize
%
%
%
%
%
%
%
%
%
%-----------------------------------------------------------------------------------------------------------------------------------
\end{document}